\documentclass[10pt,aps,prl,superscriptaddress,twocolumn]{revtex4-1}

\usepackage{amsmath}
\usepackage{amssymb}
\usepackage{graphicx}

\begin{document}

\title{The Scientific Competitiveness of Nations} 

\author{Giulio Cimini}
\email{giulio.cimini@roma1.infn.it}
\affiliation{Istituto dei Sistemi Complessi (ISC) CNR, UoS ``Sapienza'', Dipartimento di Fisica, Universit\`a Sapienza, Piazzale Aldo Moro 5, 00185 Rome, Italy}
\author{Andrea Gabrielli}
\affiliation{Istituto dei Sistemi Complessi (ISC) CNR, UoS ``Sapienza'', Dipartimento di Fisica, Universit\`a Sapienza, Piazzale Aldo Moro 5, 00185 Rome, Italy}
\affiliation{IMT Institute for Advanced Studies Lucca, Piazza San Ponziano 6, 55100 Lucca, Italy}
\author{Francesco Sylos Labini}
\affiliation{Centro Studi e Ricerche Enrico Fermi, Via Panisperna 89 A, Compendio del Viminale, 00184 Rome, Italy}
\affiliation{Istituto dei Sistemi Complessi (ISC) CNR, Via dei Taurini 19, 00185 Rome, Italy}

\begin{abstract}
We use citation data of scientific articles produced by individual nations in different scientific domains to determine the structure and efficiency of national research systems. 
We characterize the scientific fitness of each nation---that is, the competitiveness of its research system---and the complexity of each scientific domain by means of a non-linear iterative algorithm 
able to assess quantitatively the advantage of scientific diversification. We find that technological leading nations, beyond having the largest production of scientific papers 
and the largest number of citations, do not specialize in a few scientific domains. Rather, they diversify as much as possible their research system. On the other side, less developed nations 
are competitive only in scientific domains where also many other nations are present. Diversification thus represents the key element that correlates with scientific and technological competitiveness. 
A remarkable implication of this structure of the scientific competition is that the scientific domains playing the role of ``markers'' of national scientific competitiveness are those 
not necessarily of high technological requirements, but rather addressing the most ``sophisticated'' needs of the society. 
\end{abstract}

\maketitle 

\section*{Introduction}

Measuring the quality of research on national scale is of great interest to stakeholders and policy-makers for deciding on, {\em e.g.}, funding allocations and scientific priorities. 
In a seminal work, May \cite{may} analyzed the output and outcomes from research investment over years 1981-1994, comparing scientific research outputs among several nations from a variety of viewpoints. 
King \cite{king} presented a similar, more refined analysis over years 1993-2002. With the aim of determining the scientific impact of nations, King built a rank of nations ordering them 
according to their share of world citations. In addition to the national contribution to the world scientific production, the outputs and outcomes relative to the population and gross domestic product (GDP) 
were estimated. Finally, King made a comparison of five main scientific branches (medical sciences; natural sciences; agricultural sciences; engineering and technology; social sciences) 
across different nations. The citation share for each branch was then used as a branch-specific metric for quantifying the scientific impact of nations. This analysis showed, for instance, 
that Russia and Germany are relatively strong in physical science, as France is in mathematics, while United Kingdom and United States excel in medical and environmental sciences. 
The natural question arising from these studies is then whether nations tend to specialize or diversify their scientific research, and which choice is more efficient---in terms of scientific competitiveness. 
Because science is nothing but an output of the society, this issue is closely related to that of industrial production of nations, for which some of the main classical economic theories prescribe 
specialization \cite{smith,ricardo}, whereas, recent studies \cite{hidalgo,tacchella,cristelli} reveal that nations are extremely diversified and tend to produce anything they can, 
{\em i.e.}, anything compatible with their capabilities (infrastructures, technology level, education system, State efficiency, etc.). 

Here we analyze the extensive and intensive (that is, normalized to the resources invested) research performance of nations in different research sectors. 
Our goals are: (i) to determine whether the most developed nations tend to maximally diversify their research system---as for industrial production---or instead to specialize in a few scientific domains 
where their competitiveness is sufficiently high; (ii) detect the scientific sectors that are the best marker of the global scientific development and competitiveness of a nation. 
Once the importance of diversification also in scientific production is verified, in order to assess quantitatively the comparative advantage of scientific diversification we use a novel approach 
\cite{tacchella,cristelli} based on coupled non-linear maps, whose fixed point defines a metric for the scientific fitness of nations ({\em i.e.}, the competitiveness of their research systems) 
and for the complexity of scientific domains. This approach allows to identify the nations having either the more productive or efficient research system, as well as the scientific domains 
representing the best ``markers'' of the development level of national scientific research system. 

\section*{Materials and Methods}

For this analysis we use bibliometric data over years 1996-2012, collected by the Scimago website (www.scimagojr.com)---based on the Scopus database (www.scopus.com; 
note that previous studies \cite{may,king} used bibliometric data provided by Thomson Institute for Scientific Information (ISI), and are thus not directly comparable to ours). 
In particular, we use citation counts which indicate scientific impact better than the number of publications \cite{may,king}. 
The resulting citation statistics comprehend $N=238$ nations, $D=27$ scientific domains and $d=307$ scientific sub-domains (each belonging to one domain). 
Note that, given the large number of papers produced by a nation, we can safely assume that distortions that may affect a single paper are smoothed out \cite{baccini}. 
Yet, we excluded from the following analysis the multidisciplinary domain (that is difficult to classify), as well as all sub-domains with poor statistics, \emph{i.e.}, with 
a total number of citations less than a threshold $\Theta=10^3$. The filtered database then comprises data for $D=26$ domains and $d=296$ sub-domains. 
Finally note that, while in the following analysis we use data on sub-domains, we have checked that results presented in this paper are not qualitatively altered by directly using data on domains. 
Hence in what follows we use the term ``domain'' to refer to sub-domains, but the discussion applies to domains as well.
We remark that it is largely debated whether Scopus or other citation databases fairly cover all scientific domains as, for instance, Scopus collects only documents written in English 
and published in international peer-reviewed journals. In this situation the most penalized branches appear to be the ones of social sciences and humanities, 
because significant parts of the scholarly production in these areas is not published in international journals, but in national journals, in book chapters or in monographs, 
resulting in a scarce and biased coverage of the database that penalizes particularly non-anglophone nations \cite{sivertsen}. 
Nevertheless, for completeness we decided to include these domains (psychology; arts and humanities; social sciences; economics, econometric and finance; business, management and accounting) 
in our analysis---yet, our results are not particularly affected by the presence of such disciplines (as explained in the caption of Fig.\ref{Figure_list}).

\begin{figure}[t!]
\centering
\includegraphics[width=8.6cm]{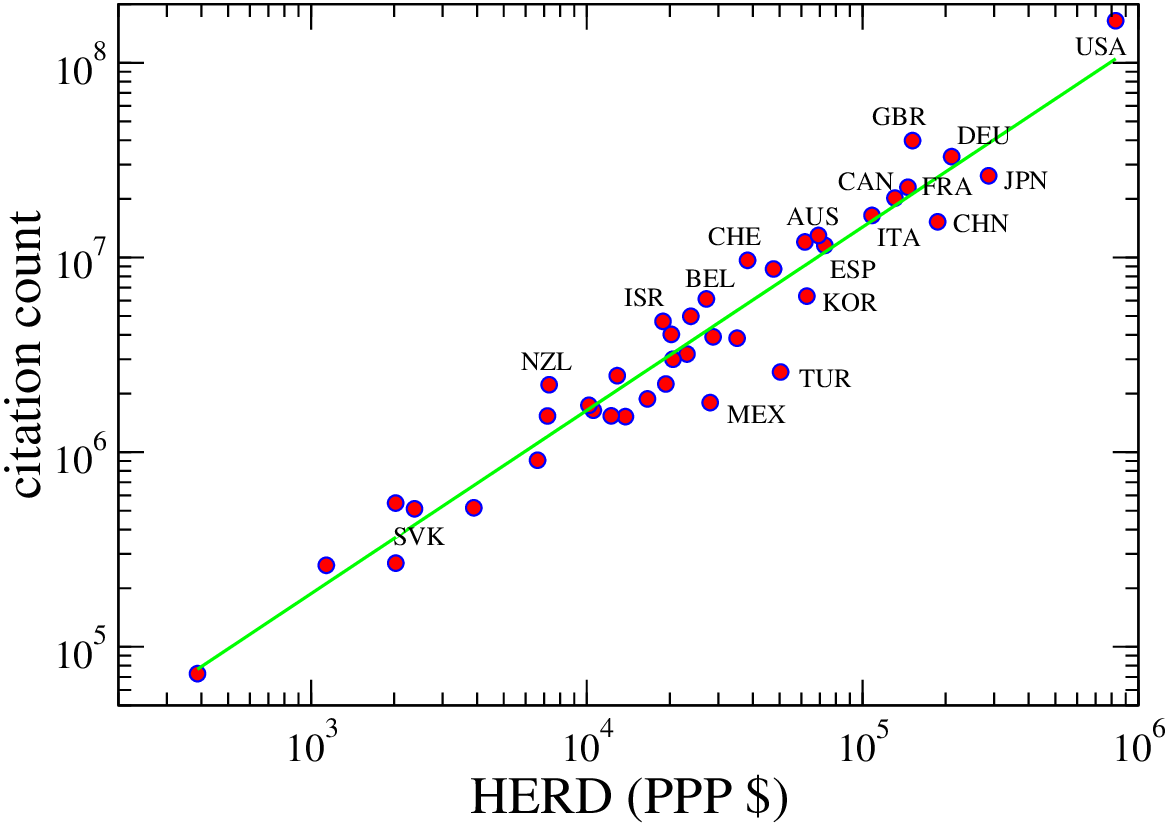}
\caption{Relation between total number of citations and HERD (expressed in PPP \$) for each nation. 
The green line is a power-law correlation of the type $\sum_a n_{ia}=b\,s_i^\gamma$, where $b$ is a constant and $s$ is the HERD. The best fit exponent 
is close to one ($\gamma=0.94\pm0.02$)---with correlation coefficient $0.97$. Such relation simply implies that the output of the scientific research, measured by the number of citations, 
depends on the resources that a nation has invested in it. The nations above/below the fit ({\em e.g.}, GBR, CHE, ISR, NZL above; JPN, CHN, MEX, TUR below) are the more/less efficient in their scientific production: 
a more refined analysis is then required to identify the structural reasons for these fluctuations.}\label{Figure_rel}
\end{figure}
We represent the dataset as a binary bipartite network, in which nodes are of two kinds: nations and scientific domains. 
Links can exist only between nations and scientific domains, and the bipartite adjacency matrix $\mathbf{M}$, with elements $m_{ia}$ equal to 1 if a link between nation $i$ and domain $a$ exists, 
and to 0 otherwise, describes the whole pattern of connections. Denoting as $n_{ia}$ the number of citations that nation $i$ has received for publications relative to domain $a$, 
in order to build the adjacency matrix $\mathbf{M}$ (and hence the network) we put $m_{ia}=1$ if the $i$-th nation ranks in the top-$T$ for number of citations in domain $a$ ({\em i.e.}, $n_{ia}$), 
and $m_{ia}=0$ otherwise. This approach is purely based on the number of citations, clearly promoting ``successful'' nations that achieved many citations---this approach is biased 
towards nation size, and thus we call the adjacency matrix built in this way as {\em extensive}. As an alternative approach, we can normalize citation counts to the actual resources 
that a nation has invested to produce those citations, namely to its expenditure in scientific research and development (R\&D). 
We quantify the latter through data of Higher Education expenditure on Research \& Development (HERD), which we obtained from the Organization for Economic Cooperation and Development 
(OECD, http://www.oecd.org/). In order to compute $s_i$ (the HERD of nation $i$ over years 1996-2012, to which citation counts refer), we use the average of the available data 
(the OECD database covers years 2000-2012) proportionally rescaled to properly quantify HERD values over years 1996-2012. 
The {\em intensive} version of the adjacency matrix is then built by putting $m_{ia}=1$ if the $i$-th nation ranks in the top-$T$ for number of citations per unit spent on HERD in domain $a$ 
({\em i.e.}, $n_{ia}/s_i$), and $m_{ia}=0$ otherwise. This alternative approach promotes ``efficient'' nations: those that---{\em ceteris paribus}---achieved more citations 
(see Figure \ref{Figure_rel}). Note that since HERD data are available only for $N=40$ developed nations, our analysis based on the intensive matrix is restricted to this subset of nations. 
We remark that both approaches to build the adjacency matrix use citations to directly compare different nations, but not different scientific domains, and are thus not biased towards domains 
with overall high number of citations. Instead, the different domains are compared only after their complexity values are evaluated---as explained below. 
Finally, let us stress that other methods to build the adjacency matrix are indeed possible. For instance, we can place a link between $i$ and $a$ whenever $n_{ia}/\sum_j n_{ja}\ge p$ (extensive) 
or $(n_{ia}/s_i)/(\sum_j n_{ja}/s_j)\ge p$ (intensive). For reasonable values of $p$, this approach does not show qualitative differences with the top-$T$ methodology.

Once the adjacency matrix of the bipartite network is built, we use the iterative algorithm proposed in Refs. \cite{tacchella} and \cite{cristelli} 
to compute the scientific fitness of the different nations and the complexity of the various scientific domains. 
This approach is motivated---and, as we will see, justified---by the approximate triangular shape or the high nestedness \cite{almeida} 
of the adjacency matrix, which shows when nations (rows) are ordered from more to less diversified, and scientific domains (columns) from more to less ubiquitous. 
These features justify the highly non-linear relationship in the algorithm to define the complexity of scientific domains from the fitness of nations making research on them, so that, as explained below, 
the complexity of a given scientific domain is mainly determined by the fitnesses of the less scientifically developed nations that are competitive in such sector. 
While this idea was originally applied to the industrial production of nations \cite{tacchella,cristelli}, it can be easily translated to the scientific research system. 
Indeed, the observation (based again on the triangularity of the matrix) that a developed nation actually produces successful research outputs in a scientific domain gives only limited information 
on the complexity of the domain itself, because this nation does research in almost all domains. On the other hand, when an underdeveloped nation is able to do research in a given domain, 
very likely this domain requires a low level of sophistication (of the discipline itself, or of the national scientific, industrial and social substrate). These observations 
lead to the following main argument behind the mathematical approach: while it is reasonable to measure the competitiveness of a nation through the sum of the complexities of the domains belonging 
to its research pool, it is not possible to adopt a similar linear approach to measure the complexity of scientific domains. A natural choice is instead to weight the fitnesses of the nations 
making research on them in a highly nonlinear way, so that a domain on which scarcely competitive nations make research achieves low complexity. On the other hand, the only possibility for 
a domain to achieve high complexity is to be part of the research system of only the highly competitive nations. The simplest way in order to translate these ideas into a quantitative measure is 
to employ a non-linear diffusive process on the bipartite network of nations and scientific domains. The iterative algorithm at the basis of such process works as follows \cite{tacchella,cristelli}. 
Denoting the fitness of nation $i$ as $f_i$ and the complexity of domain $a$ as $c_a$, the algorithm starts from evenly distributed values $f_i^{(0)}=1$ $\forall i$ and $c_a^{(0)}=1$ $\forall a$. 
Fitness and complexity values at iteration $t$ ($f_i^{(t)}$ and $c_a^{(t)}$) are then obtained from fitness and complexity values of the previous iteration ($f_i^{(t-1)}$ and $c_a^{(t-1)}$) as:
\begin{eqnarray}
\label{eq.F} \tilde{f}_i^{(t)}&=&\sum_a m_{ia}c_a^{(t-1)} \\
\label{eq.C} \tilde{c}_a^{(t)}&=&\left[\sum_i m_{ia}/f_i^{(t-1)}\right]^{-1} \;. 
\end{eqnarray}
Such values have to be normalized at the end of each iteration to obtain 
\begin{eqnarray}
\label{eq.Fg} f_i^{(t)}&=&N\,\tilde{f}_i^{(t)}/\sum_i\tilde{f}_i^{(t)} \\
\label{eq.Cg} c_a^{(t)}&=&d\,\tilde{c}_a^{(t)}/\sum_a\tilde{c}_a^{(t)} \;.
\end{eqnarray}
The algorithm is solved iteratively until fitness and complexity values converge to a fixed point:\\ 
$\sum_i |f_i^{(t)}-f_i^{(t-1)}|+\sum_a |c_a^{(t)}-c_a^{(t-1)}|<\delta$, with arbitrary small $\delta$ (here we use $\delta=10^{-12}$). 
Fitness and complexity values computed in this way are used to produce a ranking of nations and of scientific domains, which will be discussed in the next section.

\section*{Results}

\subsection*{Adjacency matrices}

\begin{figure}[t!]
\centering
\includegraphics[width=8.6cm]{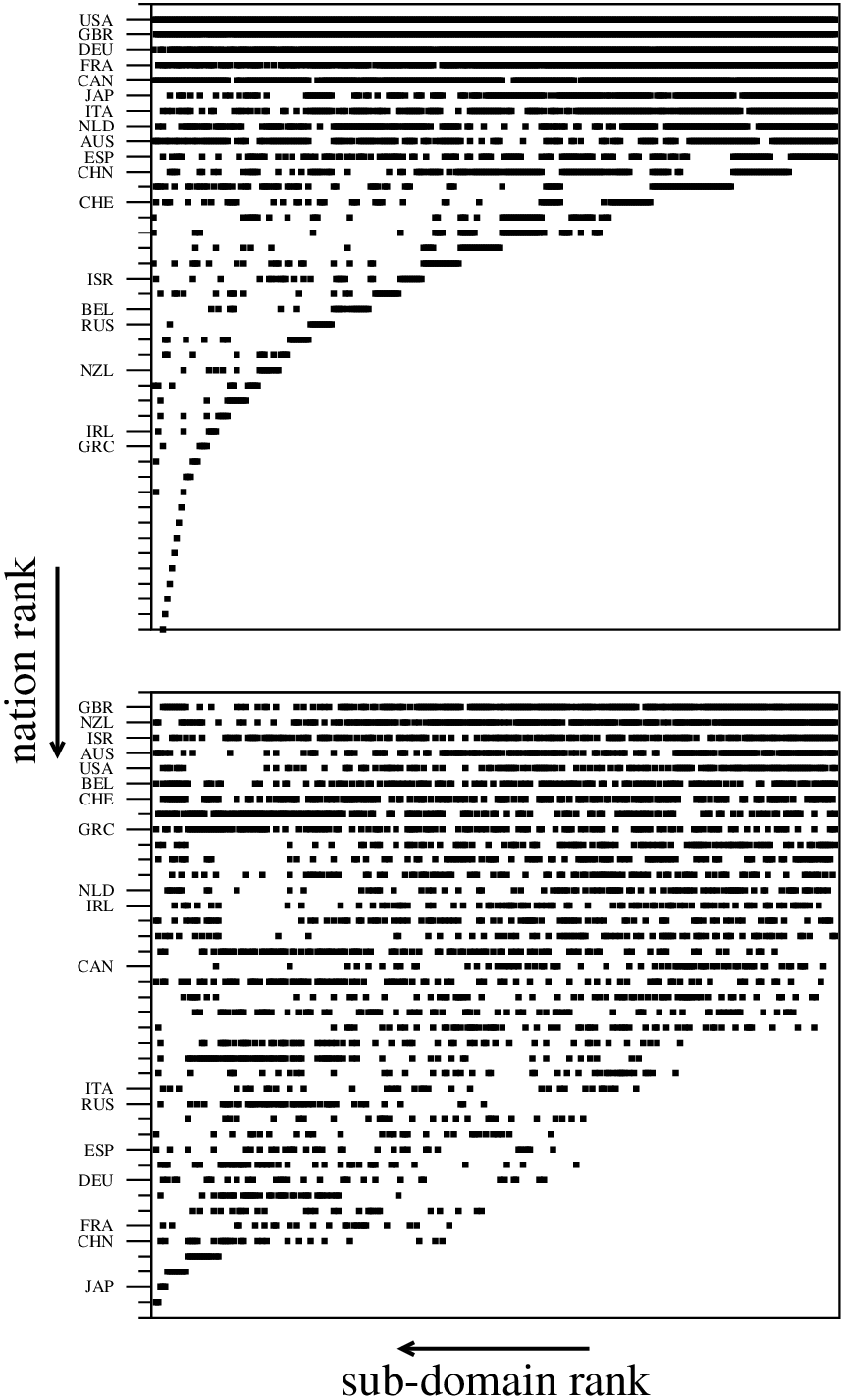}
\caption{Extensive (top panel) and intensive (bottom panel) adjacency matrices. Rows and columns are ordered according to the ranking of nations and scientific sub-domains, respectively 
(from first to last in the direction of the arrows). The labels on the vertical axes help to identify several nations in the ranking. 
The matrices were obtained for $T=10$. Indeed, the value of $T$ must be chosen not too low neither too high in order to avoid having an empty or full adjacency matrix, respectively. 
In fact, by construction the matrices have $T$ entries in each column (the top-$T$ nations in that domain), and thus a total of $NT$ entries ($N$ is the number of nations).}\label{Figure_matrix}
\end{figure}
Figure \ref{Figure_matrix} shows the extensive and intensive adjacency matrices, in which rows and columns are reordered according to nation fitnesses and scientific domain complexities 
(top nations are placed in upper rows and top domains in the far right-hand side columns). For reasonable values of $T$ (see caption of Fig. \ref{Figure_matrix}), 
in both cases we observe triangular-shaped matrices, indicating that successful nations possess an extremely diversified research system, and that such diversification decreases for less successful nations. 
Moreover less successful nations are competitive only in sectors in which many other nations are active. Note that this picture defies the standard economic approach of the wealthiest nations producing 
(making research on) only a few products (scientific domains) with high complexity, which would result in a optimal strategy only in a static situation, whereas, the strongly dynamical situation 
of the world market (science) suggests that flexibility and adaptability are mostly important to be competitive in a competition-driven changing system---in analogy with bio-systems evolving 
in a competitive dynamical environment \cite{tacchella,cristelli}. 

To quantify the triangularity of the matrices, we use the concept of nestedness---in particular, we measure the nestedness based on overlap and decreasing fill (NODF) \cite{almeida}. 
We compute the Z-score between the degree of nestedness of the observed matrices with that of random matrices from two null models: the ``semi-fixed'', 
in which matrix fill and column sums are constrained to observed values; and the ``fixed-fixed'', in which also row sums are constrained. 
Despite the NODF over columns for the observed matrices is necessarily zero (as every column has $T$ entries by construction), the Z test with the ``semi-fixed'' model returns values of 19 and 14 
for the extensive and intensive matrices, respectively, whereas, with the more constrained ``fixed-fixed'' model the test is more strict yet returns satisfactory Z values of 0.7 and 8 
for the extensive and intensive matrices, respectively---revealing a more robust triangular structure for the intensive matrix 
(note that the Z-test for a matrix with no triangular structure returns values much smaller than one). 

\subsection*{Nations ranking}

Row labels of Figure \ref{Figure_matrix} show the nations achieving the highest fitness for their scientific research system. As expected, the way of building the extensive matrix brings G8 nations 
like Unites States, United Kingdom, Germany, France and Japan to the top of the ranking. A more interesting picture emerges from the approach of the intensive matrix. 
Compared to the previous case, now only United Kingdom remains at the top of the ranking, which is now occupied by nations like Switzerland, Israel, Australia and New Zealand 
which are generally considered to be ``efficient'' (concerning money investments in research and developments). United States suffer a slight downgrade, whereas, 
all major European nations and even more far east nations (China and Japan) lose a notable number of positions. 
This points out  to the difference between having a large-scale research system and an efficient and well-designed one. 

We remark that the results presented above are confounded to a certain extent because a large and growing fraction of scientific work involves international collaborations \cite{ioannidis}, 
and because of the English language bias in the Scopus database---both in the journals included and in patterns of citation. 
The latter observation could explain to a certain degree why anglophone nations like United States, United Kingdom and Canada do much better than, {\em e.g.}, Germany, France, Italy, Japan and China.

\subsection*{Domains ranking}

\begin{figure*}[t!]
\centering
\includegraphics[width=10.75cm]{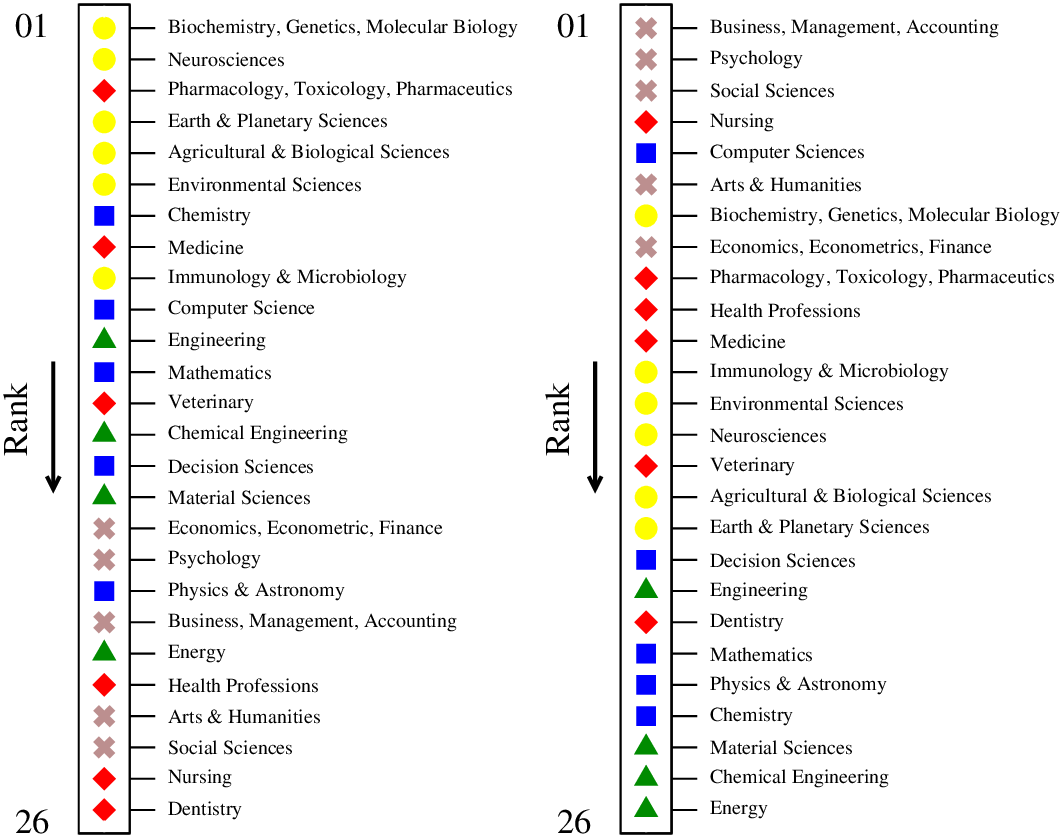}
\caption{Ranking of scientific domains from the extensive (left panel) and the intensive (right panel) matrix---from first to last in the direction of the arrows.
The ranking derives from the averages of the complexity values of the constituents sub-domains obtained over a range of $T$ values around $T=10$. 
The different symbols represent the five main branches of scientific domains: 
Yellow circles for earth and life sciences (earth and planetary sciences; environmental science; agricultural and biological sciences; biochemistry, genetics, molecular biology; neuroscience; 
immunology and microbiology); Green triangles for engineering and technology (engineering; chemical engineering; materials science; energy); 
Red diamonds for medical sciences (medicine; pharmacology, toxicology, pharmaceutics; nursing; health professions; dentistry; veterinary); 
Blue squares for physical and formal sciences (chemistry; physics and astronomy; mathematics; computer science; decision sciences); 
Brown crosses for social sciences and humanities (psychology; arts and humanities; social sciences; economics, econometric and finance; business, management and accounting). 
Note that excluding from the analysis the domains belonging to social sciences and humanities (and the associated sub-domains) leads to a ranking of nations which remains unaltered 
from what is shown in Fig.\ref{Figure_matrix}, and to to a ranking of scientific domains almost identical to what would be obtained by removing all brown crosses from the above panels. 
In this case, the rankings derived from the extensive and intensive approaches also appears more similar to each other, as generally top domains would belong to earth and life sciences, 
together with medicine and pharmacology, toxicology and pharmaceutics---with nursing and health professions being the only domains achieving a substantial upgrade in the intensive matrix approach.}\label{Figure_list}
\end{figure*}

Figure \ref{Figure_list} shows the full ranking of scientific domains. For better readability of the figure, scientific domains are divided into five main branches 
(and represented by different symbols): earth and life sciences (yellow circles), engineering and technology (green triangles), medical sciences (red diamonds), physical and formal sciences (blue squares), 
and social sciences and humanities (brown crosses). For the approach of the extensive matrix, we observe that top domains belong to life sciences, immediately followed by earth sciences. 
Medicine and especially pharmacology, toxicology and pharmaceutics also occupy top positions. Browsing the ranking down we find (in order): physical and formal sciences, engineering and technology, 
social sciences and humanities, and the other medical sciences. Some remarkable changes appear for the approach of the intensive matrix: social sciences and humanities as well as some medical sciences 
(nursing and health professions, in particular) now occupy top positions. This is probably influenced by the fact that these domains have overall a few number 
of citations and thus are more subject to noise and bias. However, this also depends on the fact that only very competitive and advanced nations develop a strong activity in these sophisticated domains. 
In other words, we can include them among the good indicators for the R\&D level of a nation. 

Despite the particular differences between the extensive and intensive approaches, it is clear from the previous analysis that top domains, {\em i.e.}, the ones that basically belong 
to the research pool of only the most competitive nations, are complex in the sense that they require a developed research system (already including all the other more fundamental domains), 
as well as a developed society. Underdeveloped nations instead are still at a stage of construction of their R\&D system, whose fundamental bricks are obviously the ``basic'' sciences 
(physical and formal) and those that are more related to economic returns (engineering and technology), which thus achieve low complexity. Under this view, the complexity of a scientific domain 
is associated not necessarily to technical requirements, but to a complex social and economic substrate that allows (and requires) making research on them. 

\section*{Discussion}

Technological leading nations have the largest production of scientific papers and collect the largest number of citations \cite{may,king}. 
They also have the highest fraction of research and development (R\&D) expense with respect to their GDP \cite{press}: indeed only nations that spend close to 3\% of their GDP in R\&D compete most successfully. 
In this study we have analyzed the scientific research system of nations, using both an extensive and an intensive ({\em i.e.}, normalizing citation counts to the resources invested) approach. 
After building a bipartite network of nations and scientific domains, we obtained values for nation scientific fitness (that is, the competitiveness of the research system) 
and scientific domain complexity as the fixed-point values of an appropriate non-linear diffusion process on the same network justified by the nested nature of the bipartite network. 
We thus showed that the adjacency matrix of the network has a triangular shape, indicating that (as for industrial production) the most successful and technological leading nations do not specialize 
in a few scientific domains---rather, they diversify as much as possible their research system. Our analysis thus points diversification out as the key element for nations to achieve a successful 
and competitive research system, suggesting that excellence comes out as a natural side effect of a complex, very heterogeneous, diversified, and therefore healthy, system. 
It is interesting to note that a recent quantitative study of the distribution of grants to research groups \cite{fortin} concluded that is more effective to award small grants 
to many researchers rather than to give large grants to a few groups of elite researchers. This conclusion is coherent with our results, namely that strategies targeting diversity, 
rather than excellence, are likely to be more effective. 

The advantage given by scientific diversification allowed us to built a ranking for the global productivity and for the efficiency of national research systems. 
Moreover, the ranking of scientific domains---based on the quantification of their complexity---reveals that more complex disciplines are not necessarily those of high technological requirements. 
Instead, they are disciplines (such as environmental conservation, medical caretaking and treatment research, and socio/economic studies) addressing needs of a more developed society 
that are not directly related to ``basic'' nor economically-driven research. In turn, these domains play the role of good ``markers'' for the scientific fitness of a nation: only highly competitive nations, 
in terms of scientific research, can have the necessary human and financial resources for the development of these disciplines.
\newline

{\bf Acknowledgments.} We thank A. Baccini, A. Banfi, G. De Nicolao and L. Pietronero for useful discussions and comments. 
This work was supported by the European project FET-Open GROWTHCOM (grant num. 611272) and the Italian PNR project CRISIS-Lab.

\end{document}